\documentclass[12pt,a4paper]{article}%

\usepackage{amsmath,amssymb,amsfonts}
\usepackage{caption2}
\usepackage[]{hyperref}
\usepackage[pdftex]{color,graphicx}
\usepackage{multicol}

\numberwithin{equation}{section} 
\setcaptionmargin{1cm}
\newcommand{\hhref}[1]{\href{http://arxiv.org/abs/#1}{{\it arXiv:#1}}}
\begin{document}
\begin{titlepage}
$\quad$
\vskip 2.0cm
\begin{center}
{\Large \bf Supersymmetry without a light Higgs boson} \\
\vskip 0.1 cm
{\Large \bf but with a light pseudoscalar}
\vskip 1.0cm {\large  Paolo Lodone} \\[1cm]
{\it Scuola Normale Superiore and INFN, Piazza dei Cavalieri 7, 56126 Pisa, Italy} \\[5mm]
\vskip 1.0cm
\end{center}

\begin{abstract}
We consider the $\lambda$SUSY model, a version of the NMSSM with large $\lambda H_1 H_2 S$ coupling, relaxing the approximation of large singlet mass and negligible mixing of the scalar singlet with the scalar doublets.
We show that there are regions of the parameter space in which the lightest pseudoscalar can be relatively light, with unusual consequences on the decay pattern of the CP-even Higgs bosons and thus on the LHC phenomenology.
\end{abstract}
\end{titlepage}

\section{Introduction }

The hierarchy problem, or the theoretical difficulty in accommodating for a small Higgs boson mass with respect to the ultraviolet (UV) completion of the Standard Model (SM), provides the best hope to see new physics at the CERN LHC.
In the case of the Minimal Supersymmetric Standard Model (MSSM) or its non-minimal extensions, one expects many new particles to show up in the range of hundreds of GeV at most.
First of all, however, one has to look for the (lightest) Higgs boson himself, and the presence of these new light particles can result in a significant distortion in the Higgs boson couplings and decay channels with respect to the SM.

This is of interest for the LHC not only in order to estimate the actual discovery potential, but also because in some situations the main decay channels can be totally non-standard.
The former motivation amounts to the observation that the LHC at 7-8 TeV with $\sim 1$ fb$^{-1}$ of integrated luminosity is expected to be able to exclude at 95\% c.l. a SM-like Higgs boson with mass up to about 500 GeV \cite{talkCamporesi}. Thus it makes sense to study configurations with non-standard Higgs decay channels in motivated models, so that we can know which model would be able to survive a negative result from the direct searches.
The latter motivation is more urgent, since one has to know whether it is sufficient to look for a SM-like Higgs boson and then eventually properly rescale the cross section, or instead the experimental collaborations may have to look also for different signatures.

A well-known potentially problematic possibility is a Higgs boson decaying in two light pseudoscalars $h\rightarrow aa$, where each $a$ decays predominantly in $b\overline{b}$ (or 2 jets if $m_a < 2m_b$), $\tau^+ \tau^-$, and eventually two neutralinos, giving rise to final states with large SM background at the LHC. The importance of this decay channel was first emphasized in \cite{Gunion:1996fb}, later in \cite{Dobrescu:2000jt,Dobrescu:2000yn}, and then extensively studied in \cite{Ellwanger:2001iw,Ellwanger:2003jt,Ellwanger:2004gz,Ellwanger:2005uu}. In these last studies it is investigated whether it is possible to establish, for the Next to Minimal Supersymmetric Standard Model (NMSSM), a `no-loose theorem' stating that at least one Higgs boson should be discovered via the usual SM-like production and decay channels at the LHC throughout the entire parameter space. The result is that it is not easy to cover all the possibilities: in some difficoult cases 300 fb$^{-1}$ of integrated luminosity at 14 TeV of c.o.m. energy are needed for a $5\sigma$ discovery, in other cases the theorem is not established yet.
Models with the presence of a light pseudoscalar have been discussed also recently, both in supersymmetric \cite{Franceschini:2010qz,Gunion:2011hs} and non-supersymmetric \cite{Gripaios:2009pe} contexts.
Moreover recent results go in the direction of a complete proof of the no-loose theorem, such as \cite{Mahmoudi:2010xp}, \cite{Kaplan:2011vf} or \cite{Almarashi:2010jm,Almarashi:2011bf,Almarashi:2011te,Almarashi:2011hj}, although further study is still needed.

This paper is focused on the case of $\lambda$SUSY, that is a version of the NMSSM with relatively large $\lambda H_1 H_2 S$ coupling \cite{Barbieri:2006bg}. This model has received recent attention because it naturally allows a supersymmetric spectrum of non-standard type \cite{Barbieri:2010pd}\footnote{See also \cite{Lodone:2010st}.} in which the supersymmetric flavour and CP problems are ameliorated by means of hierarchical sfermion masses \cite{Dine:1990jd,Dine:1993np,Pouliot:1993zm,Pomarol:1995xc,Barbieri:1995uv,Cohen:1996vb,Barbieri:1997tu}\footnote{See also \cite{Giudice:2008uk} for a recent study.}, realizing to some extent a `more minimal'-like configuration \cite{Cohen:1996vb} without a large amount of finetuning \cite{Barbieri:1987fn,Dimopoulos:1995mi}.
This is possible thanks to the increase in the lightest Higgs boson mass at tree level, which can be as large as 250-350 GeV.\footnote{See also \cite{Lodone:2010kt} for a comparison with other models with increased $m_h^{tree}$.} The price to pay is that the coupling $\lambda$ becomes large at a relatively low scale, signalling a change of regime in the theory which is left unspecified from a bottom-up point of view. As shown in the early studies of this idea, also called `supersymmetric fat Higgs' \cite{Harnik:2003rs,Chang:2004db,Delgado:2005fq,Birkedal:2004zx}\footnote{See also \cite{Gherghetta:2011wc} for a recent proposal.}, it is remarkable that possible UV completions do exist and can even be compatible with gauge coupling unification. The details of the UV completion are however irrelevant for the TeV-scale phenomenology, whose features are mainly determined by the large superpotential coupling $\lambda$. 

The model under discussion has been thoroughly studied in the case of relatively large singlet mass, and thus small singlet vacuum expectation value (vev) and mixing with the other states \cite{Barbieri:2006bg,Cavicchia:2007dp}.
The main purpose of this paper is to slightly extend the study of $\lambda$SUSY to the case of lighter singlet with sizable vev and mixing with the scalar doublets, paying attention in particular to unusual Higgs decay channels.
We show that there are regions of the parameter space in which the decay channels $h \rightarrow aa$ and $h \rightarrow a Z$ are dominant, where $h$ is a CP-even Higgs boson and $a$ is a CP-odd one, thus reinforcing the need of dedicated experimental studies in order to extract as much as possible from the SM background.

The model is defined in Section \ref{sect:model} and the relevant parameter space in in Section \ref{sect:bounds}.
Interesting `non-typical' configurations are then discussed in Section \ref{sect:results}.
For a recent study of the NMSSM along similar lines, but in the complementary `scale invariant' version of the model without superpotential mass terms, see also \cite{Bertuzzo:2011ij}.

\section{$\lambda$SUSY} \label{sect:model}

The model is defined by the superpotential\footnote{With this convention on the sign of $\lambda$ we have a positive $v_s$ for positive $\lambda$ and large positive $M_S$, see (\ref{eq:vevS}). Notice in fact that everything is invariant under $S \, (v_s) , \lambda \rightarrow -S \, (-v_s), -\lambda$.\label{foot1}}:
\begin{equation}
 W = \left( \mu - \lambda S \right) H_1 H_2 + \frac{1}{2} M_S S^2 
\end{equation}
with large-ish $\lambda$, and we consider the case of zero A-term ($A_\lambda=0$) and no cubic term $\frac{k}{3}S^3$ (i.e. $k=0$, $A_k=0$). 
Introducing an explicit $\mu$ term means that we give up a solution of the $\mu$ problem, focussing on the Higgs mass and finetuning problems. This approach is similar to \cite{Delgado:2010uj}.

Since the dominant effects come from the large coupling $\lambda$, we can neglect for simplicity the gauge couplings in the scalar potential and assume large gaugino masses. The result is:
\begin{equation}
V = \mu_1^2(S) |H_1|^2 + \mu_2^2(S) |H_2|^2 - (\mu_3^2(S) H_1 H_2 + h.c.) + \lambda^2 |H_1 H_2|^2 + V_S(S) \, .
\end{equation}
with $H_1 H_2 = H_1^0 H_2^0 - H_1^- H_2^+$.
In this approximation $\lambda \gg g,g'\rightarrow 0$ and no tadpole term we have, including the usual soft masses $m^2_{H_{1,2}}$, the soft term $b$ related to $\mu$, and the soft mass of the singlet $m_S^2$:
\begin{eqnarray}
\mu_{1,2}^2(S) &=& (|\mu|^2 + m_{H_{1,2}}^2) - (\lambda \mu S + h.c.) + |\lambda|^2 |S|^2 \\
\mu_3^2(S) &=& b + \lambda M_S S  \quad , \quad
 V_S(S) = (|M_S|^2 + m_S^2)|S|^2  \label{eq:mu3} 
\end{eqnarray}
Notice that the `effective' $\mu-$term is given by:
\begin{equation}
\mu_{eff} = \mu - \lambda v_s \,
\end{equation}
where $v_s$ is the vev of the singlet scalar.
For simplicity we will assume all the various parameters to be real.
The minimization gives:
\begin{eqnarray}
\frac{v_2}{v_1} &=& \tan \beta = \frac{\mu_1(v_s)}{\mu_2(v_s)} \label{eq:minimForv1} \\
\lambda^2 v^2 &=& m_A^2 - \mu_1^2(v_s) - \mu_2^2(v_s) \qquad (\mbox{where } m_A^2 = \frac{2 \mu_3^2(v_s)}{\sin 2\beta}) \\
0 &=& \mu_1^{2 \prime}(v_s) v_1^2 + \mu_2^{2 \prime}(v_s) v_2^2 - 2 \mu_3^{2 \prime}(v_s) v_1 v_2 + V_S^{\prime}(v_s) \, \label{eq:mimimForS}
\end{eqnarray}
where $\sqrt{v_1^2 + v_2^2} = v = 174$ GeV is the usual electroweak vev.
From the last equation we obtain, in our case:
\begin{equation} \label{eq:vevS}
\quad v_s =  
 \frac{\lambda v^2 (\mu_{eff} + \frac{1}{2} M_S  \sin 2\beta)}{\mu_S^2} \, .
\end{equation}
We then see that for large $\mu_S^2 = |M_S|^2 + m_S^2$ the vev $v_s$ is small.
What we want to do is to partially extend the analysis of \cite{Barbieri:2006bg}\cite{Cavicchia:2007dp} to the case of smaller $\mu_S$, and thus sizable $v_s$ and mixings.

For our purposes it is sufficient to compute the various scalar masses at tree level.
Let us consider the CP-even scalar mass matrix, assuming no spontaneous CP violation as imposed in Section \ref{sect:bounds}. We define:
\begin{equation}
Re[H_1^0] = v_1 + \frac{S_1}{\sqrt{2}} \quad , \quad Re[H_2^0] = v_2 + \frac{S_2}{\sqrt{2}} \quad , \quad Re[S] = v_s + \frac{S_3}{\sqrt{2}}
\end{equation}
and the mass matrix in the $(S_1,S_2,S_3)$ basis takes the form ($v = 175$ GeV):
\begin{equation} \label{eq:evenMassMatrix2}
M^{(+)} = 
\left(
\begin{array}{ccc}
m_A^2 \sin^2 \beta & (\lambda^2 v^2 - \frac{1}{2} m_A^2)\sin 2\beta & -2\lambda \mu_{eff} \, v \, \cos\beta - \lambda M_S v \sin\beta \\
 & m_A^2 \cos^2 \beta &  -2\lambda \mu_{eff}\, v \, \sin\beta - \lambda M_S v \cos\beta \\
 &  & \mu_S^2 + \lambda^2 v^2
\end{array}
\right)\,  .
\end{equation}
For the CP-odd scalar sector we have:
\begin{equation}
Im[H_1^0] =  \frac{P_1 \sin \beta - G^0 \cos\beta}{\sqrt{2}} \quad , \quad Im[H_2^0] =  \frac{P_1 \cos\beta + G^0 \sin\beta}{\sqrt{2}} \quad , \quad Im[S] =  \frac{P_2}{\sqrt{2}}
\end{equation}
where $G^0$ is the degree of freedom that is eaten-up to give mass to the weak vectors.
In the basis $(P_1,P_2)$ the mass matrix is:
\begin{equation} \label{eq:oddMassMatrix2}
M^{(-)} = 
\left(
\begin{array}{cc}
m_A^2  & \lambda M_S v  \\
 &  \mu_S^2 + \lambda^2 v^2
\end{array}
\right)\, .
\end{equation}
Finally in this `gaugeless' approximation we have: $m_{H^{\pm}}^2 = m_A^2 - \lambda^2 v^2$.

\section{Parameters and constraints} \label{sect:bounds}

A full set of free parameters is given by:
\begin{equation}
\lambda \, , \,  \tan\beta \,  , \, m_{H^{\pm} } \, , \, \mu_{eff} \, , \, M_S \, , \, \mu_S \,  \, .
\end{equation}
There are then some theoretical constraints on the parameter space.
First of all we have to require that the potential is stable. 
A sufficient condition is:
\begin{equation}  \label{eq:bound0}
\mu_1^2(0) \, , \, \mu_2^2(0) >0
\quad \Leftrightarrow \quad
m_{H^{\pm}}^2 \cos^2 \beta + 2 \lambda \mu_{eff} v_s + \lambda^2 v_s^2 >0 \,
\end{equation}
\begin{equation}  \label{eq:bound1}
|\mu|^2 < \mu_1^2(0) \, , \, \mu_2^2(0)
\quad \Leftrightarrow \quad
m_{H^{\pm}}^2 > \frac{\mu_{eff}^2}{\cos^2 \beta} \,
\end{equation}
besides $\mu_S^2 >0$. It is also clear that under these conditions the charged fields do not take a vev, so the electromagnetism is unbroken.
Notice that, after imposing the above conditions, without special relations among the parameters there are only two stationary points: $v_1 = v_2 = v_s=0$ and the solution in which all the three vevs are nonzero.
Thus the analysis in \cite{Kanehata:2011ei} is not of concern in this case.

The condition for EWSB is then:
\begin{equation} \label{eq:bound2}
\mu_1^2(v_s) \mu_2^2(v_s) < (\mu_3^2(v_s))^2 \quad \Leftarrow \quad m_{H^\pm}^2 >0 \, .
\end{equation}
We should now impose that the Hessian is positive definite in $(v_1,v_2,v_s)$, so that this point is a local minimum. Hovever since the potential is stable and the only other stationary point is $(0,0,0)$, it is sufficient to impose $V(v_1,v_2,v_s) < V(0,0,0)=0$, that amounts to:
\begin{equation} \label{eq:bound3}
v_s^2 \mu_S^2 < \frac{\lambda^2 v^4}{4}\sin^2 2\beta
\quad \Leftrightarrow \quad
|\mu_{eff} + \frac{\sin 2\beta}{2} M_S | < \frac{\sin 2\beta}{2} \mu_S \, 
\end{equation}
with $\mu_S >0 $ (see also below).
We also set $|\mu_{eff}|>100$ GeV to satisfy the LEP bound on the chargino masses, in fact in our limit of heavy gaugino masses the lightest chargino has mass $|\mu_{eff}|$.

Finally we impose that there is no spontaneous CP violation. Replacing $v_2 \rightarrow v_2 e^{i\theta} \, , \, v_s \rightarrow v_s e^{i\phi}$, we obtain in the potential the following phase-dependent terms:
$$
\Delta V = -2 \lambda \mu v_s v^2 \cos\phi - 2 \mu_3^2(0) v_1 v_2 \cos\theta - 2 \lambda M_S v_s v_1 v_2 \cos(\theta + \phi) \, .
$$
The condition for avoiding spontaneous CP violation, with a potential of the type $V=A \cos \phi + B \cos \theta + C \cos(\phi+\theta)$, can be written in the form $ A + C >0$ and $B + AC + BC >0$.
This is equivalent to impose that the eigenvalues of (\ref{eq:oddMassMatrix2}) are real and positive.
The resulting condition produces cuts on the values of $\mu_{eff}$, independently of $m_{H^\pm}$.
Notice that we are not assuming $v_s>0$ (see footnote \ref{foot1}).


Let us finally discuss in which portion of the allowed parameter space we can have a light pseudoscalar.
Looking at the the CP-odd scalar mass matrix (\ref{eq:oddMassMatrix2}), we see that to have a small eigenvalue we need $M_S \gtrsim \mu_S$, which implies a negative soft mass $m_S^2 <0$. Because of (\ref{eq:bound3}) we also need $\mu_S$ not much different from $|\mu_{eff}|$, and $\mu_{eff}<0$.



\begin{figure}
\begin{center}
\begin{tabular}{|c|c|c|c|c|c|}
\hline
  $\mu_S=1000$  &  $M_S=500$  &  $\tan\beta= 1.5$  & $\lambda=2$ &  $\mu_{eff}=120$  &  $m_{H^{\pm}}=300$ \\
\hline \hline
  $m_h=192$  & $m_{H_1}=338$ &  $m_{H_2}= 1088$ & $m_{A_1}=424$  &  $m_{A_2}= 1075$ & $m_{\chi_1}=63$    \\
\hline
 $G_{hVV}=0.91$  &  $G_{htt}=1.16$ &  $\Gamma_{h}^{tot}=5.0$  &  $G_{H_1 VV}=0.36$  &  $G_{H_1 tt}=0.26$  &  $\Gamma_{H_1}^{tot}=1.7$
\\
\hline \hline
decay channel: & $ ZZ$ & $ WW$ & $ A_1 Z$ & $ \chi\chi$ & $ A_1 A_1$ \\
\hline
$BR_h$ (\%) & 4.7 & 14.0 & 0 & 81.3 & 0 \\
\hline
$BR_{H_1}$ (\%) & 30.9 & 67.3 & 0 & 1.8 & 0 \\ \hline
\end{tabular}
\\ \vspace{0.5cm}
\begin{tabular}{cc}
\includegraphics[width=0.38\textwidth]{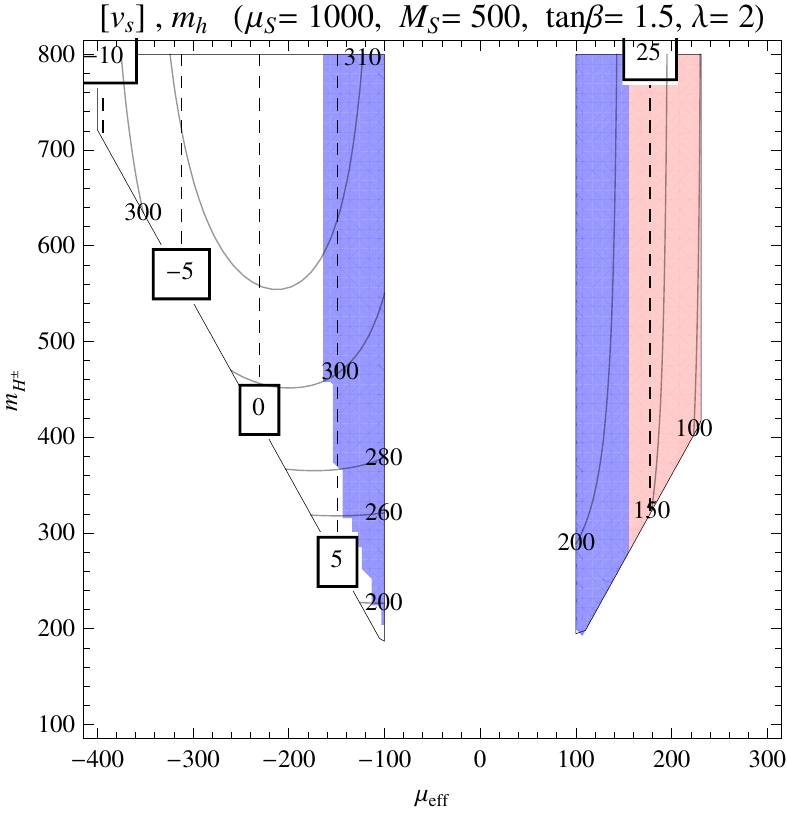} &
\includegraphics[width=0.40\textwidth]{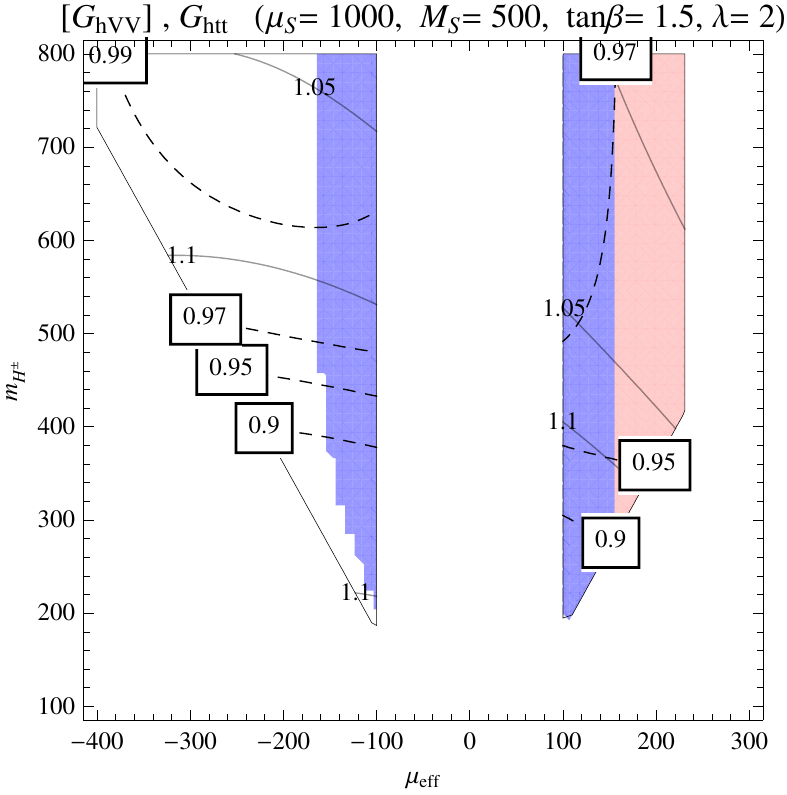}\\
\includegraphics[width=0.39\textwidth]{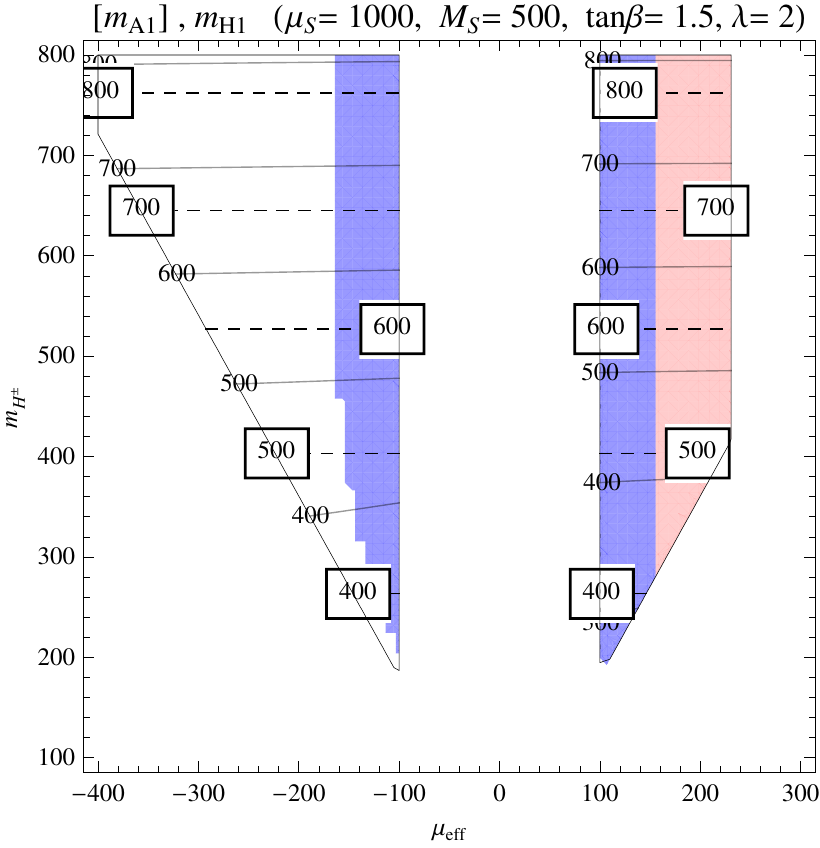} &
\includegraphics[width=0.39\textwidth]{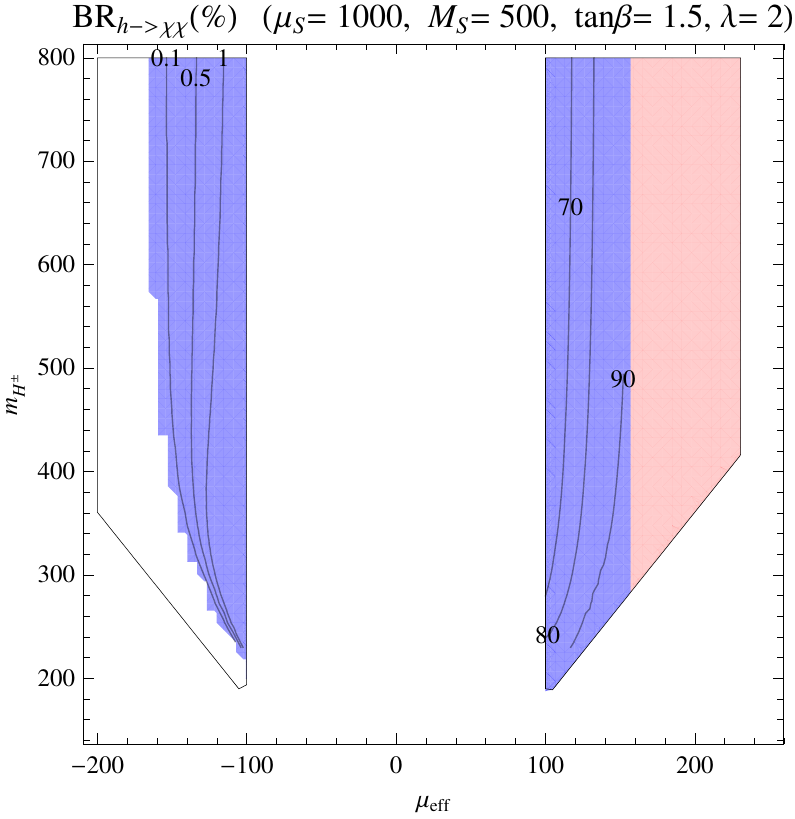}
\end{tabular}
\caption{{\small \it Contourplots of the relevant masses, couplings and Branching Ratio (BR) in the $(\mu_{eff},m_{H^{\pm}})$-plane, in a configuration with heavy pseudoscalar, for comparison. The quantity in squared brackets in the caption is shown with dashed lines and framed labels. The dark shaded region corresponds to $m_h > 2 m_{\chi}$, the light shaded one is excluded because $m_\chi < m_Z /2$. A sample point is given in the Table above. Masses in GeV.}}
\label{fig:standardcase}
\end{center}
\end{figure}

\begin{figure}
\begin{center}
\begin{tabular}{|c|c|c|c|c|c|}
\hline
  $\mu_S=300$  &  $M_S=600$  &  $\tan\beta= 1.5$  & $\lambda=2$ &  $\mu_{eff}=-150$  &  $m_{H^{\pm}}=300$ \\
\hline \hline
  $m_h=233$  & $m_{H_1}=256$ &  $m_{H_2}= 552$ & $m_{A_1}=50$  &  $m_{A_2}= 650$ & $m_{\chi_1}=145$    \\
\hline
 $G_{hVV}=0.84$  &  $G_{htt}=0.76$ &  $\Gamma_{h}^{tot}=4.5$  &  $G_{H_1 VV}=0.32$  &  $G_{H_1 tt}=0.92$  &  $\Gamma_{H_1}^{tot}=2.3$ \\
\hline \hline
decay channel: & $ ZZ$ & $ WW$ & $ A_1 Z$ & $ \chi\chi$ & $ A_1 A_1$ \\
\hline
$BR_h$ (\%) & 13.5 & 32.0 & 0.3 & 0 & 54.2 \\   
\hline
$BR_{H_1}$ (\%) & 6.1 & 14.2 & 55.2 & 0 & 24.5 \\  
\hline
\end{tabular}
\\ \vspace{0.5cm}
\begin{tabular}{cc}
\includegraphics[width=0.39\textwidth]{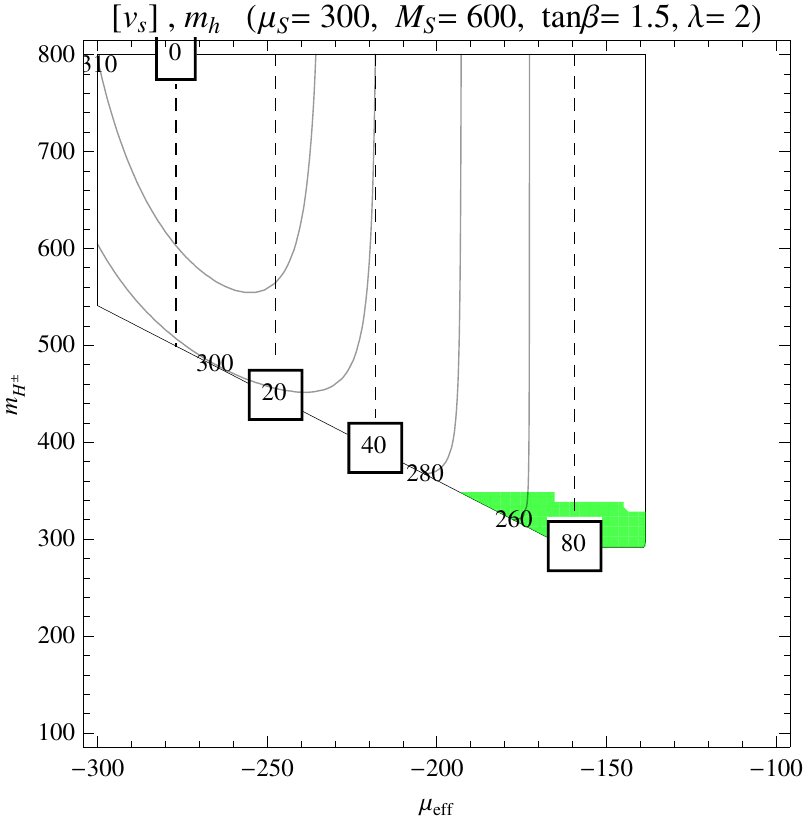} &
\includegraphics[width=0.39\textwidth]{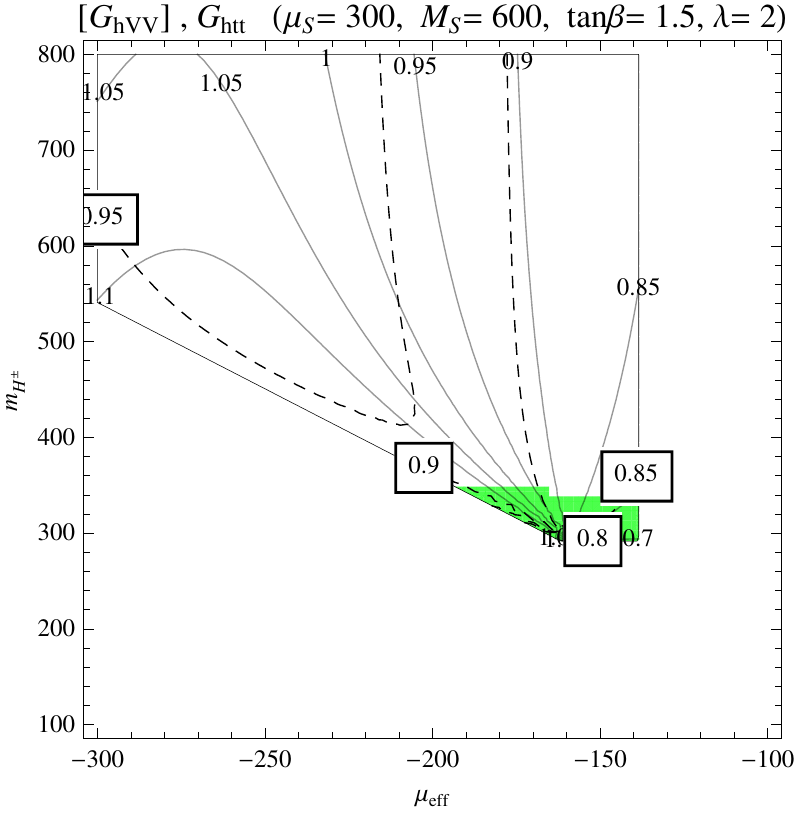}\\
\includegraphics[width=0.39\textwidth]{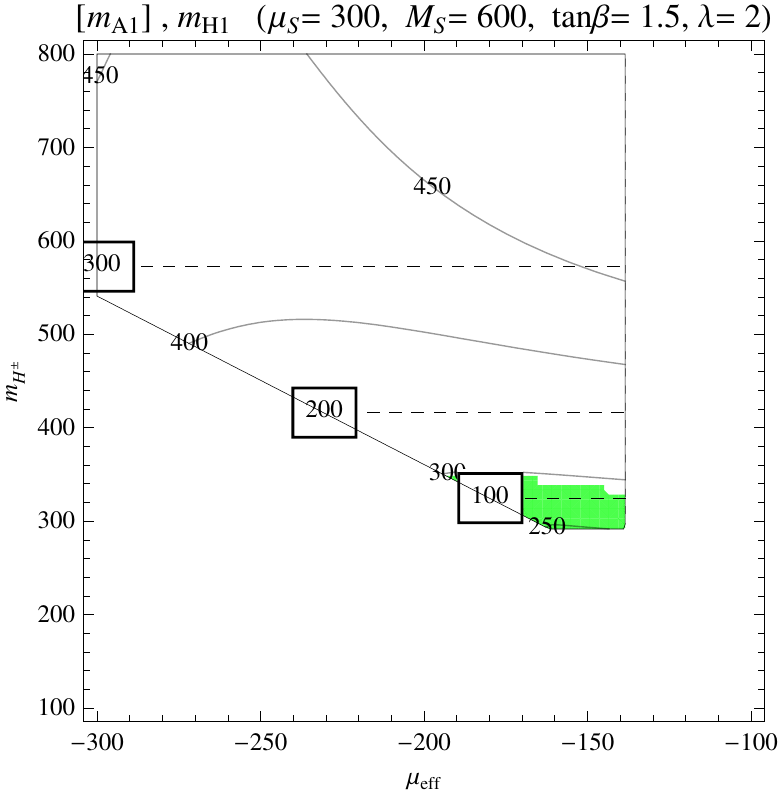} &
\includegraphics[width=0.39\textwidth]{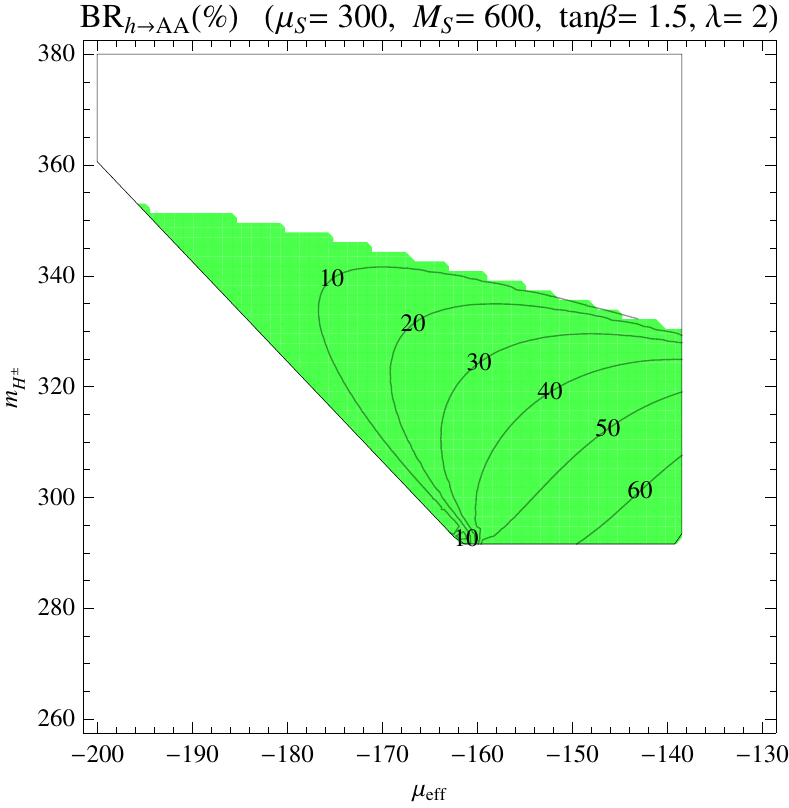}
\end{tabular}
\caption{{\small \it Contourplots of the relevant masses, couplings and BR in the $(\mu_{eff},m_{H^{\pm}})$-plane, in a configuration with a light pseudoscalar. The quantity in squared brackets in the caption is shown with dashed lines and framed labels. The shaded region corresponds to $m_h > 2 m_{A_1}$. A sample point is given in the Table above. Masses in GeV.}}
\label{fig:relevantcase}
\end{center}
\end{figure}

\section{Results and Conclusions} \label{sect:results}

We are mostly interested in studying the production and decays of the lighter neutral CP-even Higgs bosons $h$ and $H_1$, which can mimic a SM-like Higgs boson with mass below 500 GeV.
The dominant production mechanism for the neutral scalars is the gluon fusion, as long as the coupling to the top quark is not extremely suppressed \cite{Mahmoudi:2010xp}\cite{Cavicchia:2007dp}.
We can thus quickly estimate the production cross section for $s=h,H_1$ by simply multiplying the production cross section of the SM Higgs boson times the square of the reduced coupling $G_{s tt}=g_{stt} / g_{stt}^{SM}$, which depends only on the mixing angles (see Appendix, or Table 3 of \cite{Mahmoudi:2010xp}).
This estimate is precise in the limit of relatively heavy stops, as naturally allowed in this context, so that their contribution to the gluon fusion amplitude is subdominant.
For the decay the main modes are, when kinematically allowed: $h,H_1 \rightarrow WW$, $ZZ$, $A_1 Z$, $A_1 A_1$, $\chi\chi$, where $\chi$ is the lightest neutralino, and: $H_1 \rightarrow hh$, $t\overline{t}$. We neglect for simplicity all the other subdominant modes, since the LHC projection of 95\% exclusion of a Higgs boson below about 500 GeV \cite{talkCamporesi} is mostly based on the decay into vectors.
This also means that the actual discovery potential can be estimated by rescaling the SM Higgs boson production cross section 
by $G_{htt}^2 \times BR(h \rightarrow WW,ZZ)$, and analogously for $H_1$ whenever $m_{H_1}<500$ GeV.

In the Figures \ref{fig:standardcase} and \ref{fig:relevantcase} we report two representative examples of what can happen, focussing on the situation where $m_{H_1} < 2m_h , 2m_t$ (for the case in which these channels are open too, see \cite{Cavicchia:2007dp}).
In the first one we report, for comparison, a `typical $\lambda$SUSY' configuration with heavy pseudoscalar and inverted spectrum $m_h < m_{H^{\pm}} \lesssim m_{H_1} < m_{A_1}$. In this case the only channel that can deplete the decay of the light scalars into vectors is the decay in two neutralinos.
With our conventions (see footnote \ref{foot1}) we have a heavier $h$ with smaller coupling to $\chi\chi$ for $\mu_{eff}<0$ - small $v_s$, and a lighter $h$ with larger coupling to $\chi\chi$ for $\mu_{eff}>0$ - sizable $v_s$.
In the case of Figure \ref{fig:relevantcase} we have a `non-typical' $\lambda$SUSY configuration with $A_1$ sufficiently light 
 so that the channels $h\rightarrow A_1 A_1$ and $h\rightarrow A_1 Z$ are open, and we see that the first one can be dominant in some regions of the parameter space.

\begin{figure}
\begin{center}
\begin{tabular}{cc}
\includegraphics[width=0.39\textwidth]{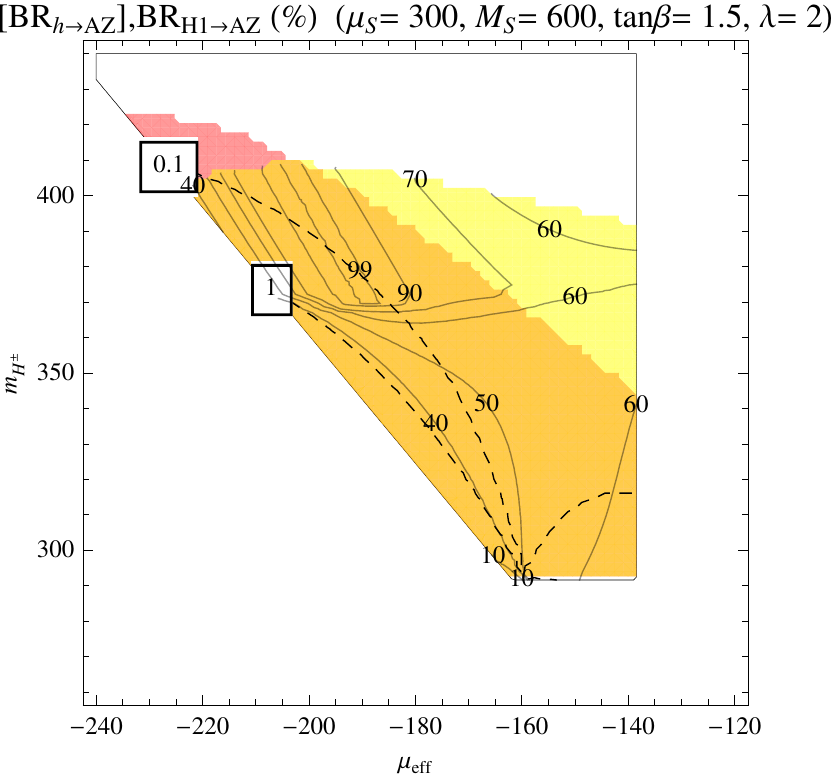} &
\includegraphics[width=0.39\textwidth]{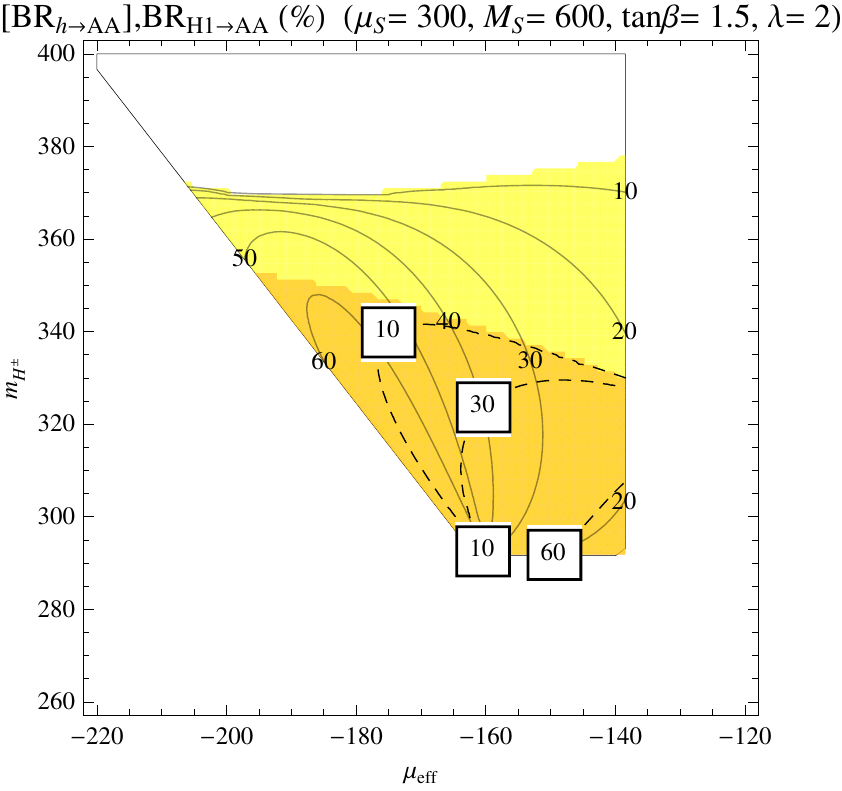}
\end{tabular}
\caption{\small {\it Contourplots of the BRs of $h$ and $H_1$ in $A_1 Z$ and $A_1 A_1$ in the same configuration as in Figure \ref{fig:relevantcase}, in the $(\mu_{eff},m_{H^{\pm}})$-plane. The quantity in squared brackets in the caption is shown with dashed lines and framed labels. In the shaded region the corresponding channel is open, and $m_{H_1}<2 m_t , 2m_h$ so that the channels $H_1 \rightarrow hh, \,t\overline{t}$ are closed. Masses in GeV.}}
\label{fig:relevantBR}
\end{center}
\end{figure}

Figure \ref{fig:relevantBR} gives a more precise idea of what the $BR$s look like, including those of $H_1$. We see that for both $h$ and $H_1$ we can have $BR(\rightarrow A_1 A_1) \sim 60\%$, thus depleting again the decay into vectors. For the decay in $A_1 Z$ the situation can be even more peculiar, especially where the $H_1\rightarrow A_1 A_1,\chi\chi$ modes are closed. In fact in this region $H_1$  can decay only in $A_1 Z$ or vectors; but the two corresponding couplings are orthogonal (see Appendix), so that there is a small region in which it decays practically  only\footnote{Remembering that we neglected other subdominant channels.} in $A_1 Z$.

To sum up, the conclusion of this quick analysis of `non-typical $\lambda$SUSY' is duplex.
First of all we demonstrated that there are configurations in which the coupling of the `candidate SM-like Higgs bosons' to the top and their $BR$s into vectors are significantly reduced. The relevant cross sections can be reduced by a factor of order $10$, thus allowing the model to survive an eventual negative result from the first direct searches at the LHC although with smaller allowed  parameter space.
More importantly, we showed that there are cases in which unusual modes like $A_1 Z$ and $A_1 A_1$, with $A_1$ going then predominantly in $b\overline{b}$ and $\tau \overline{\tau}$, can dominate the decays of one or both of $h$ and $H_1$. 
This indicates that it may be necessary to further study these unusual channels in order to try to extract as much as possible from the SM background, that is not easy to tame.

It is remarkable for example that the possibility \cite{Dermisek:2005gg,Dermisek:2006wr,Dermisek:2007yt} that a Higgs boson decaying mainly as $h\rightarrow aa \rightarrow 4\tau$ had escaped the LEP search has been ruled out\footnote{Although there may still be room for this possibility if $a$ decays mainly into hadrons \cite{Domingo:2011rn}.} by the ALEPH Collaboration only in recent times, by re-analyzing data many years after the closing of LEP \cite{Schael:2010aw}. This shows how important is that all the non-standard possibilities are kept in mind from the very beginning, since most likely at the LHC, due to the strong selection in the recorded events, such later analyses will not be possible at all.

\section*{Acknowledgments}

I thank Riccardo Barbieri for useful suggestions.
This work is supported in part by the European Programme ``Unification in the LHC Era",  contract PITN-GA-2009-237920 (UNILHC).


\section*{Appendix: Relevant couplings}

For completeness, we report the couplings that are relevant for the above discussion.
We define $g_{P_1 P_2 P_3}$ as what enters in the Feynman rule for ${P_1, P_2, P_3}$, apart from a factor $i$ and eventually the metric $g_{\mu\nu}$ or terms like $(p_1 + p_2)^\mu \epsilon_\mu$.

 For the couplings $hVV$ and $htt$ we define the reduced couplings (see also Table 3 of \cite{Mahmoudi:2010xp}):
\begin{equation}
G_{hVV} = \frac{g_{hVV}}{g_{hVV}^{SM}} \quad , \quad G_{htt} = \frac{g_{htt}}{g_{htt}^{SM}} \, ,
\quad \mbox{ where: } \quad
g_{hVV}^{SM} = \sqrt{2} \frac{m_V^2}{v} \quad , \quad g_{htt}^{SM} = \frac{m_t}{v \sqrt{2}}\, \, .
\end{equation}
If the mass matrix (\ref{eq:evenMassMatrix2}) is diagonalized by the rotation: $R^{(+)T} M^{(+)} R^{(+)}$, and we call the mass eigenstates as:
$
(S_1,S_2,S_3)^T = R^{(+)} (h,H_1,H_2)^T \, ,
$ 
we then have:
\begin{equation}
G_{hVV} = R^{(+)}_{11} \cos \beta + R^{(+)}_{21} \sin \beta \,\, \stackrel{2HDM}{\rightarrow} \sin (\beta - \alpha) \quad , \quad G_{htt} = \frac{R^{(+)}_{21}}{\sin\beta} \,\, \stackrel{2HDM}{\rightarrow} \frac{\cos\alpha}{\sin\beta} \, .
\end{equation}
where in the last step we recovered the result of a general 2 Higgs Doublet Model in the limit of no mixing with $S_3$, with $R^{(+)}_{11}= -\sin \alpha$ and $R^{(+)}_{21}=\cos\alpha$.
Here and below, the corresponding couplings of $H_1$ are given by the same expression with the replacement: $R^{(+)}_{x1} \rightarrow R^{(+)}_{x2}$.
%

 For the coupling $h A_1 Z$ we have (see also \cite{Miller:2003ay}):
\begin{equation}
\mathcal{L} = \frac{\sqrt{g^2 + g^{\prime 2}}}{2} Z^\mu \left( S \partial_\mu P_1 + P_1 \partial_\mu S \right)
\qquad \mbox{ where: } \qquad
S=-\sin\beta\, S_1 + \cos\beta\, S_2 \, .
\end{equation}
As a result, if we diagonalize the mass matrix (\ref{eq:oddMassMatrix2}) through the rotation $R^{(-)T} M^{(-)} R^{(-)}$, and we define:
$
(P_1,P_2)^T = R^{(-)} (A_1,A_2)^T \, ,
$
then the Lagrangian coupling (or, equivalently, what multiplies $i(p_h + p_{A_1})_\mu \epsilon_Z^\mu$ in the Feynman rule) is given by:
\begin{equation}
g_{hA_1 Z} = \frac{\sqrt{g^2 + g^{\prime 2}}}{2} (-\sin\beta\, R^{(+)}_{11}  + \cos\beta\, R^{(+)}_{21} )R^{(-)}_{11} \quad \stackrel{2HDM}{\rightarrow} \frac{\sqrt{g^2 + g^{\prime 2}}}{2} \cos(\beta -\alpha) \, .
\end{equation}
Notice that this combination of scalar fields is orthogonal to the one with coupling to $VV$.
%

 The Lagrangian for the coupling $h A_1 A_1$ is:
\begin{eqnarray}
\mathcal{L} &=& \frac{\lambda \mu_{eff}}{\sqrt{2}} S_3 P_1^2 -\frac{\lambda^2 v}{\sqrt{2}}P_2^2 (S_1 \cos\beta + S_2 \sin\beta) \\
&& -\frac{\lambda M_S}{\sqrt{2}} \left[ S_3 P_1^2 \frac{\sin 2\beta}{2} + P_1 P_2 \left( S_1 \cos\beta + S_2 \sin\beta \right) \right] - \frac{\lambda^2 v}{\sqrt{2}} \left[ S_1 \cos^3 \beta + S_2 \sin^3 \beta \right]P_1^2 \, , \nonumber
\end{eqnarray}
so that the corresponding coupling is given by:
\begin{eqnarray}
&\frac{g_{hA_1 A_1}}{2} = \frac{\lambda \mu_{eff}}{\sqrt{2}} R^{(+)}_{31} (R_{11}^{(-)})^2 -\frac{\lambda^2 v}{\sqrt{2}}(R_{21}^{(-)})^2 (R^{(+)}_{11} \cos\beta + R^{(+)}_{21} \sin\beta) 
-\frac{\lambda M_S}{\sqrt{2}} \left[ R^{(+)}_{31} (R_{11}^{(-)})^2 \frac{\sin 2\beta}{2} \right.& \nonumber \\
& \left. + R_{11}^{(-)} R_{21}^{(-)} \left( R^{(+)}_{11} \cos\beta + R^{(+)}_{21} \sin\beta \right) \right] 
 - \frac{\lambda^2 v}{\sqrt{2}} \left[ R^{(+)}_{11} \cos^3 \beta + R^{(+)}_{21} \sin^3 \beta \right](R_{11}^{(-)})^2  \, . &
\end{eqnarray}
%

 For the decay in two neutralinos we consider the limit in which the gauginos are heavy or equivalently the limit of small gauge couplings, so that the singlino mixes with the Higgsinos only.
The relevant Lagrangian is then:
\begin{eqnarray*}
&\mathcal{L} 
=  -\frac{1}{2}\left( (\mu - \lambda S) (-\tilde{N}_1 \tilde{N}_1 + \tilde{N}_2 \tilde{N}_2) - \lambda H_1 (\tilde{S} \frac{\tilde{N}_2 - \tilde{N}_1}{\sqrt{2}} + \leftrightarrow) - \lambda H_2 (\tilde{S} \frac{\tilde{N}_2 + \tilde{N}_1}{\sqrt{2}} + \leftrightarrow )  +  M_S  \tilde{S}\tilde{S}  + h.c. \right)&
\end{eqnarray*}
where $\tilde{N}_1 = \frac{\tilde{H}_1-\tilde{H_2}}{\sqrt{2}}$ , $\tilde{N}_2 = \frac{\tilde{H}_1+\tilde{H_2}}{\sqrt{2}}$,
and $M_S$ is the superpotential mass term of the singlet.
The resulting mass matrix is:
\begin{equation}
\mathcal{L} = -\frac{1}{2} (\tilde{N}_1, \tilde{N}_2,\tilde{S}) \,\, M_{\chi}  \,\, (\tilde{N}_1, \tilde{N}_2,\tilde{S})^T + h.c
 \mbox{, with: } \,
M_{\chi} = 
\left(
\begin{array}{ccc}
-\mu_{eff} & 0 & \lambda \frac{v_1 - v_2}{\sqrt{2}} \\
0 & \mu_{eff} & -\lambda \frac{v_2 + v_1}{\sqrt{2}} \\
\lambda \frac{v_1 - v_2}{\sqrt{2}} & -\lambda \frac{v_2 + v_1}{\sqrt{2}} & M_S
\end{array}
\right) \, .
\end{equation}
Thus in the limit $\tan \beta \approx 1$ the combination $\tilde{N}_1=(\tilde{H}_1 - \tilde{H}_2)/\sqrt{2}$ has mass $\sim |\mu_{eff}|$, while the orthogonal combination mixes with $\tilde{S}$.
The relevant coupling is then given by:
\begin{equation}
g_{h\chi\chi} = -{\lambda}\left\{ R^{(+)}_{31}[(R^{(\chi)}_{21})^2 - (R^{(\chi)}_{11})^2] + \sqrt{2} R^{(+)}_{11} [R^{(\chi)}_{21}-R^{(\chi)}_{11}]R^{(\chi)}_{31} + \sqrt{2} R^{(+)}_{21} [R^{(\chi)}_{21}+R^{(\chi)}_{11}]R^{(\chi)}_{31} \right\}  
\end{equation}
where $M_\chi$ has been diagonalized by $R^{(\chi) T} M_\chi R^{(\chi)}$, so that:
$
(\tilde{N}_1 , \tilde{N}_2 , \tilde{S})^T = R^{(\chi)} (\chi_1 , \chi_2 , \chi_3)^T \, .
$


\vspace{0.3cm}

\begin{multicols}{2}

\end{multicols}

\end{document}